\documentclass[aps,prl,twocolumn,showpacs]{revtex4}

\usepackage{graphicx}

\begin{document}

\bibliographystyle{apsrev}

\title{Information and measurement in generally covariant quantum theory}

\author{S. Jay Olson}
 \email{solson2@lsu.edu}
 \author{Jonathan P. Dowling}
\affiliation{Hearne Institute for Theoretical Physics, Department of Physics and Astronomy, Louisiana State University, Baton Rouge, LA 70803}

\date{\today}

\begin{abstract}
Due to the absence of an external, classical time variable, the probabilistic predictions of covariant quantum theory are ambiguous when multiple measurements are considered.  Here, we introduce an information theoretic framework to the covariant formalism, and use it to interpret the measurement process.  We find that the time ordering of measurements emerges as an entropy relationship in the state of the observers, giving unique probabilities for multiple measurements.  This approach suggests a new, fully self-contained probability interpretation for generally covariant quantum physics, which makes use of a quantum mechanical description of the observer, in contrast to standard quantum mechanics which assumes an external, classical observer.
\end{abstract}


\maketitle

To formulate quantum theory in a fully background-free manner (general enough to encompass quantum cosmology, quantum gravity, etc \---- theories in which the concept of time evolution can be obscure), Reisenberger and Rovelli~\cite{Rovelli1} constructed a general quantum formalism defined for arbitrary systems, including dynamical systems without a classical background causal structure (e.g. a time parameter, or background metric).  Although this formalism clarifies the issue of dynamics in covariant quantum theories, serious issues remain with the interpretation of projective measurement in this context.  Since the formulation of the theory does not depend on a background causal structure, there is no pre-defined sense of before and after with which to order projections and obtain multiple-measurement probabilities.

To see how this becomes a serious problem, recall that in standard quantum theory, if we are given a state $|\Psi \rangle$, the probability to observe this system in state $|\phi \rangle$ is given by $| \Pi_{\phi} |\Psi \rangle |^2$, where $\Pi_{\phi}$ is the projector onto the state $|\phi \rangle$.  What is the probability that we observe the system in state $|\phi \rangle$ \emph{and} in state $|\sigma \rangle$?  It is $| \Pi_{\sigma} \Pi_{\phi} |\Psi \rangle |^2$ if we perform the $\phi$-measurement first, and $| \Pi_{\phi} \Pi_{\sigma} |\Psi \rangle |^2$ if we perform the $\sigma$-measurement first.  Since the projectors may not commute, we must specify time ordering with respect to some causal structure to obtain unique predictions.  But this kind of structure is exactly what covariant quantum theory lacks.  How are we to obtain unique probabilities?

In the original work, it was suggested that this order of projections ambiguity might be resolved in one of two ways:  If a sequence of measurements is treated as a single measurement by a larger system, the problem might be avoided.  Alternatively, it might be solved if the observer defines a time variable, and the order of projections is taken with respect to the observer's time.  These ideas have since been elaborated in more detail in Ref.~\cite{Rovelli4}.

Our analysis follows an information-theoretic description of measurement introduced by Cerf and Adami~\cite{Adami1, Adami2}, which obtains probabilities from the reduced density operator describing the observer's state.  In this formalism, sequences of measurements are analyzed without time-ordered projections \---- time ordering is recovered via an entropy relationship satisfied by the observers.  Suitably generalized to the covariant formalism, we find that this neatly solves the covariant time ambiguity.

In what follows, we begin by reviewing Cerf and Adami's description of measurement in the context of standard Schr\"odinger picture quantum mechanics (which is not covariant and makes use of a classical time variable).  We then review covariant quantum theory, and generalize the appropriate information-theoretic tools to present a prescription for obtaining unique probabilities in the covariant context.

We consider a sequence of measurements of incompatable observables.  Consider a quantum system we wish to study, $Q$, in the following state:
\begin{eqnarray}
	|Q \rangle = \sum_{i} \alpha_{i} |a_{i} \rangle
\end{eqnarray}
Performing the first measurement, $Q$ interacts with an observer system $A$ (Alice), with basis states $|i \rangle $ (which could also be read as $|\mathrm{ see\ outcome}\  i \rangle$).  The interaction is such that we obtain the following state of $QA$:
\begin{eqnarray}
	|QA \rangle = \sum_{i} \alpha_{i}|a_{i},i \rangle
\end{eqnarray}
Now we introduce a second observer system $B$ (Bob), with basis states $|j \rangle $, which will be entangled with $Q$ in a new basis $|b_{j} \rangle$ \---- eigenstates of an observable that does not commute with that measured by $A$ (i.e. the overlap $U_{ij} = \langle b_{j}|a_{i} \rangle$ is not the identity matrix).  After interaction with $B$, the system is in the following state:
\begin{eqnarray}
	|QAB \rangle = \sum_{i,j} \alpha_{i} U_{ij} |b_{j},i,j \rangle
\end{eqnarray}
This is a pure state with zero entropy.  However, we can trace over the relevant subsystems to obtain reduced density operators directly describing the observers $A$ and $B$:
\begin{eqnarray}
\rho_{A} &=& \sum_{i} |\alpha_{i}|^{2} |i \rangle \langle i|
	\\
\rho_{B} &=& \sum_{i,j} |\alpha_{i}|^{2} |U_{ij}|^{2} |j \rangle \langle j|
\end{eqnarray}
These density operators have von Neumann entropy identical to the Shannon entropy associated with classical random variables with probabilities $p_{A}(i) = |\alpha_{i}|^{2}$ and $p_{B}(j) = \sum_{i} |\alpha_{i}|^{2} |U_{ij}|^{2}$, and these are exactly the probabilities associated with the assumption of a projective collapse during the $A$-measurement, and subsequent $B$-measurement, though we have introduced no such concept.  We are free to read off the probabilities without bringing non-unitarity into the theory.  In essence, this is a decoherence formalism without an external environment.  Cerf and Adami simply noticed that every observer is decoherent with respect to the system $Q$ on which they perform measurements \---- this fact is already built into the von Neumann description of measurement.  While tracing over the system $Q$ is somewhat counterintuitive, this is exactly what must be done to obtain the information directly available to the observers.   

Since the emergent classical probabilities in this process take the form $p_{B}(j) = \sum_{i} |U_{ij}|^{2} p_{A}(i)$, it is clear that any subsequent observers will obtain from their observation process an entropy at least as great as the entropy of the preceding observers (given the pure state process above).  This is closely related to the well-known result that projective measurements can only increase the entropy of a system~\cite{Nielsen1}.  In this unitary picture, however, it is the entropy of the observer systems that is increasing rather than the entropy of the full state, which remains constant.  Due to this property, the entropy of the observers can be used to tell us the order in which measurements occurred, and we see that the full quantum mechanical arrow of time is subtly hidden in unitary Schr\"odinger picture quantum mechanics, when information theory is accounted for.  The important point is that information on the order of measurements can be recovered from the entropy of the observers themselves, even without access to the external, classical time variable.

Let us see graphically how this measurement formalism is already more general than projective measurement.  First we introduce an entanglement diagram to describe the measurement process.  In these diagrams, the central bold line represents system $Q$.  Observer systems $A$ and $B$ are represented as lines running parallel to $Q$, along the flow of time.  Interactions are represented as ``photons" running between systems.  The state of the system at any time is given by the type of line representing the observer systems:  A dashed line means the system is separable with respect to $Q$, while a solid line means the system is entangled with respect to $Q$ \---- i.e. tracing out over $Q$ leaves a mixed state of the observer system, signaling the onset of classical probabilities.
\begin{figure}
	\centering
		\includegraphics[width=0.25\textwidth]{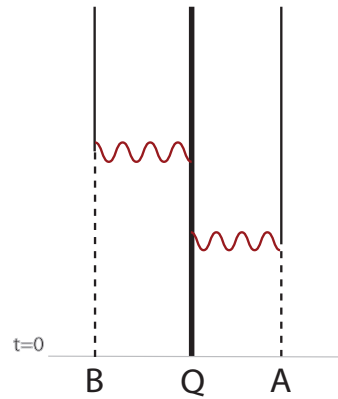}
	\caption{Measurement of two incompatable observables as in equations (1) through (5).  Time moves up the diagram.  Dashed lines indicate seperability with respect to $Q$, solid lines represent entanglement with respect to $Q$.}
\end{figure}

Using this scheme, the measurement process described by equations 1\--5 is depicted in figure 1.  Entanglement is increasing with time, so it is not surprising that the resulting entropy chain reproduces the standard theory.  However, consider figure 2.  We have deliberately drawn this diagram on its side to make the arrow of time more ambiguous.  Note that if we consider the state of this system at $t=0$ (the center), we could interpret the interactions as measurements proceeding \emph{in either direction of time}.  Evolving to the right from $t=0$, $A$ is measured first and then $B$; evolving to the left we have $B$ measured first, and then $A$.  Note that the entropic time ordering of the right-hand measurements will be different than the left-hand measurements.  If we were to arbitrarily choose one direction as that of increasing time, then time ordered projections would give us correct probabilities for one set of measurements, but wrong probabilities for the other.  While this example might appear somewhat pathological, there is no reason that such a solution to the Schr\"odinger equation could not exist, and we use it to emphasize that the entropic time ordering is more fundamental than externally time-ordered projections.
\begin{figure}
	\centering
		\includegraphics[width=0.40\textwidth]{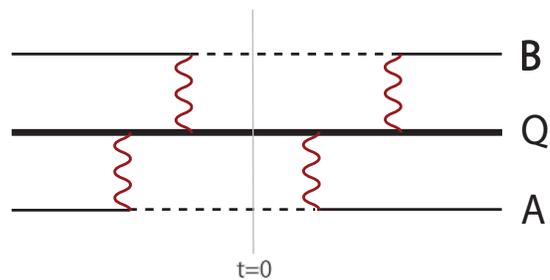}
	\caption{Measurement of $A$ and $B$ proceeding in either time direction (left or right) from initial time.  The external time variable cannot be used to consistently order all four measurements \---- instead, the entropy of the observers determines the order of the effective projections.}
\end{figure}

Moving to the covariant formalism of Reisenberger and Rovelli, we have different notions of state and evolution.  A state in the physical Hilbert space $\mathcal H$ is not an $L^{2}$ function on space, but a solution to the equations of motion (expressed as a Wheeler-DeWitt equation $H \Psi = 0$) having support throughout all spacetime (extended configuration space).  States do not evolve from one to another \---- evolution is built into the definition of physical states.

However, to \emph{specify} a physical state, we make use of a kinematical Hilbert space, $\mathcal K$ which are $L^{2}$ functions on the extended configuration space, $\mathcal M$.  These kinematical states represent local information specified by an observer \---- e.g., if a particle is known to be in a small spacetime region $\mathcal R$, we specify a $\mathcal K$ state $|R \rangle$ such that the function $R(x) = \langle x | R \rangle $ has support in $\mathcal R$~\cite{Rovelli1, Rovelli2, Rovelli3}.

Once a $\mathcal K$ state is specified, we obtain the full physical state from it by means of the projector $P$:

\begin{eqnarray}
P:\ \mathcal K & \rightarrow & \mathcal H \\
P:\	\psi^{\mathcal K}(x) & \mapsto & \psi^{\mathcal H}(x) \\
\psi^{\mathcal H}(x) & = & \int_{\mathcal M} dx' \ W(x;x') \psi^{\mathcal K}(x')  
\end{eqnarray}
where $W(x;x')$ is the propagator for the theory.  To get physical predictions from the theory, it was postulated in Ref.~\cite{Rovelli1} that the probability for a system described by $\mathcal K$ state $|\Phi \rangle$ to be observed in a small region $\mathcal R$ is given by:

\begin{eqnarray}
	\mathcal P_{R} & = & | \langle R |P| \Phi \rangle |^{2} \\
	& = &  \left| \int dx\ dx'\ R^{\ast}(x) W(x;x') \Phi (x') \right|^{2} 
\end{eqnarray}
where $R(x)$ is a uniformly smeared function over the region $\mathcal R$.  In Refs.~\cite{Rovelli1, Rovelli2}, models of measurement were constructed that support this interpretation, provided that $\mathcal R$ is sufficiently small (outside of the small-$\mathcal R$ limit, it does not reproduce the Born interpretation of the wave function~\cite{Olson2}).  Note that this postulate works only for a single measurement.  If more measurements are involved, we need to identify a time coordinate on $\mathcal M$, or a classical background causal structure to know in what order to collapse the physical state.  This violates the spirit and generality of the formalism, so we introduce here a formalism analogous to that of Cerf and Adami, which obtains probabilities without time ordered projections.

To gain intuition for how this is to be done, note that the entanglement diagrams above can each be thought of as a single (pure) state in $\mathcal H$ \---- a full solution throughout the history of a multi-component system.  However, there are many ways to express these as states in $\mathcal K$.  In practice (in the Schr\"odinger picture), we use this freedom to express an $\mathcal H$ state as $\mathcal K$ states on different constant time slices, via the restriction $\Psi^{\mathcal K}_{t=0}(X)= \Psi^{\mathcal H}(X,0)$.  Thus what we were doing above amounted to taking partial traces and computing entropies from a particular $\mathcal K$ state representation of the full physical state in $\mathcal H$, which gave us the probabilities for the observers at a particular time.

We now define a covariant system analogous to $Q$ by specifying its extended configuration space $\mathcal M_{Q}$ and a relativistic Hamiltonian $H_{Q}$.  To include a measuring system, we enlarge the configuration space via the Cartesian product, i.e. $\mathcal M = \mathcal M_{Q} \times \mathcal M_{A}$, and define a new Hamiltonian $H$ for the combined system.  Let $x$ represent coordinates of $\mathcal M_{Q}$, and let $y$ represent coordinates of $\mathcal M_{A}$.  

To study information held by subsystems, we need a partial trace.  This is a non-trivial thing on the physical Hilbert space of solutions, since $\mathcal H$ is not generally a tensor product of subsystems.  By analogy with Cerf and Adami above, we wish the partial trace to express local information held by a specific subsystem, but not by selecting a preferred time variable.  Instead, we select a generic region of interest $\mathcal S$ somewhere in $\mathcal M$.  In the limit of the Schr\"odinger picture, $\mathcal S$ corresponds to a constant time slice of space, but in general $\mathcal S$ may be chosen freely, provided a few conditions are met:  We require that the physical state $\phi^{\mathcal H}$ under consideration can be expressed via the projection of a $\mathcal K$ state $\phi^{\mathcal K}$ with support in $\mathcal S$.  Next, we require that for the points $\left\{x, y \right\}$ and  $\left\{x', y' \right\}$ in $\mathcal S$, the propagator takes the form $W(x,y;x',y') = W_{Q}(x;x')\delta(y - y')$, where $W_{Q}(x;x')$ is obtained from the free Hamiltonian $H_{Q}$.  This expresses the fact that we are considering a region where no interactions between $Q$ and $A$ are taking place and the evolution of $A$ is trivial \---- i.e. a region where $A$ has already made its transitions and is simply holding information.

When these conditions are met, we can approximate the full Wheeler-DeWitt equation $H(x,y)\Psi(x,y)=0$ as $H_{Q}(x)\Psi(x)=0$ in our region of interest $\mathcal S$, and thus the physical state space $\mathcal H$ is locally indistinguishable from $\mathcal H_{Q} \otimes L^{2}(\mathcal M_{A}) \equiv \mathcal H_{Q} \otimes \mathcal K_{A}$.

We define a new projector, $P_{Q}$, from $\mathcal K_{Q}$ to $\mathcal H_{Q}$, so that we can now express our state as a physical density operator via $\rho = P_{Q} |\phi^{\mathcal K} \rangle \langle \phi^{\mathcal K}| P_{Q}^{\dagger}$ (valid only in the region $\mathcal S$).  Now express $|\phi^{\mathcal K} \rangle$ via a Schmidt decomposition as $\sum_{i} \lambda_{i} |\phi^{\mathcal K_{Q}}_{i} \rangle |\phi^{\mathcal K_{A}}_{i} \rangle$.  Now $P_{Q}$ operates only on $\mathcal K_{Q}$, so we can take a partial trace over $\mathcal K_{Q}$ and using the cyclic property of the trace, the properties of the projector and the physical inner product ($\langle x| P^{\dagger} P| x' \rangle = W(x;x')$~\cite{Rovelli3}), we obtain a reduced density operator on $\mathcal K_{A} \equiv L^{2}({\mathcal M_{A}) }$:
\begin{widetext}
\begin{eqnarray} 
	\rho_{A} = N^{-1} \sum_{i,j} \int_{\mathcal M_{Q}} dx\ dx'\ \lambda_{i} \lambda_{j} \phi^{\mathcal K_{Q}}_{i}(x) W(x;x') \phi^{\ast \mathcal K_{Q}}_{j}(x') |\phi^{\mathcal K_{A}}_{i} \rangle \langle \phi^{\mathcal K_{A}}_{j}|
\end{eqnarray}
\end{widetext}
where $N = \int_{\mathcal M} dx\ dx' \phi^{\ast \mathcal K}(x)W(x;x')\phi^{\mathcal K}(x')$ is for normalization.  This reduced density operator on $\mathcal K_{A}$ contains the physically relevant information locally available to an observer within $\mathcal S$.  The range of integration is contained entirely within $\mathcal S$ due to the support of the $L^{2}$ functions $\phi^{\mathcal K}$.  This covariant definition immediately reduces to the standard definition of the partial trace when the region of interest $\mathcal S$ is a constant time slice, but it is clearly more general \---- $\phi^{\mathcal K}$ may be smeared in any number of ways over a non-zero time interval, provided that the system $A$ is making no transitions.  It also remains meaningful for covariant systems having no pre-defined time variable at all.

With this partial trace, we describe an idealized measurement process in analogy with Cerf and Adami.  The setup for two measurements can be described as follows:  We have two measurement systems whose configurations are described by $\mathcal M_{A}$ and $\mathcal M_{B}$, while the system under observation is described on configuration space $\mathcal M_{Q}$.

The description of a covariant collapse, then, consists in specifying three regions of the extended configuration space, $\mathcal S$, $\mathcal S'$, and $\mathcal S''$.  With respect to the above partial trace, the reduced density operators $\rho_{A}$ and $\rho_{B}$ are pure states when expressed in the region $\mathcal S$.  In region $\mathcal S'$, $\rho_{B}$ is pure, but $\rho_{A}$ can be expressed as some mixed state $\rho_{A} = \sum_{i} |c_{i}|^{2} |\phi^{\mathcal K_{A}}_{i} \rangle \langle \phi^{\mathcal K_{A}}_{i}|$.  In region $\mathcal S''$, we have that:
\begin{eqnarray}
	\rho_{A} & = & \sum_{i} |c_{i}|^{2} |\phi^{\mathcal K_{A}}_{i} \rangle \langle \phi^{\mathcal K_{A}}_{i}|\\
	\rho_{B} & = & \sum_{i,j} |c_{i}|^{2} |U_{ij}|^{2} |\phi^{\mathcal K_{B}}_{j} \rangle \langle \phi^{\mathcal K_{B}}_{j}|
\end{eqnarray}
where each detector state $\phi^{\mathcal K}_{i}$ reflects a given state (say $|R_{i} \rangle$) of the system $Q$ as before, and $U_{ij} = \langle R_{j}| P_{Q} | R_{i} \rangle$ are the transition amplitudes between these.  This structure exactly mimics the entanglement and entropy structure leading to the effective collapse above, and if we compare the entropy of these reduced density operators to the Shannon entropy of a classical distribution, we are led to the classical probabilities $p_{A}(i) = |c_{i}|^{2}$ and $p_{B}(j) = \sum_{i} |c_{i}|^{2} |U_{ij}|^{2}$.  Thus we have identified a general type of collapse in the covariant formalism which reduces immediately to the standard Schr\"odinger picture form in the appropriate limit (when $\mathcal S$, $\mathcal S'$, $\mathcal S''$ are constant-time slices), but whose general form requires no classical background causal structure or preferred configuration variable to play any special role.  As before, an entropic time ordering is contained in the observer systems themselves.

Let us review the emergent picture of this approach:  Cerf and Adami identify the information theoretic structure leading to an effective collapse.  We introduce the covariant notion of partial trace and local entropy of a subsystem defined in a particular region of the extended configuration space.  With these constructs, we express the information theoretic structure needed to identify an effective collapse without recourse to an external time variable.  The resulting entropies define their own effective time ordering in the observers.  Just as in the example of fig. 2, however, this entropic time ordering cannot in general be associated with the increase of any single configuration variable \---- it is a property of the covariant state itself.

This approach represents an alternative probability interpretation for all of covariant quantum physics.  We propose that covariant quantum states \emph{do not} encode information about probabilities for non-unitary collapse with respect to an external, classical observer (as assumed by standard QM, and the Reisenberger-Rovelli postulate).  Instead, we propose that covariant states encode probabilistic information between quantum subsystems, within specified regions of $\mathcal M$.  Failing to make this distinction is the source of the time ordering/probability ambiguities, as well as the source of trouble in recovering the Born interpretation outside of the small-$\mathcal R$ limit.  We believe that the recent progress in Ref.~\cite{Rovelli4} is a result of moving toward this paradigm.  In Ref.~\cite{Rovelli4}, the original Reisenberger-Rovelli probability postulate is maintained, but a quantum description of $N-1$ measuring systems is included to remove the time ordering ambiguity for $N$ measurements.  However, the Born correspondence problem still persists outside of the small-$\mathcal R$ limit.  Moving to our proposed interpretation solves both problems at once \---- since our prescription is by construction identical to the Cerf-Adami formalism in the flat-spacetime Schr\"odinger equation limit, there are no problems recovering the standard Born wave-function interpretation~\cite{Olson2}, and we have seen here that quantum collapse and time ordering are emergent.  

We acknowledge useful discussions with Christoph Adami, as well as support from the Disruptive Technologies Office and the Army Research Office.

\bibliography{ref}

\end{document}